\documentclass[final,12pt]{elsarticle}
\pdfoutput=1

\usepackage{slashbox}
\usepackage{amsmath}
\usepackage{amssymb}
\usepackage{amsfonts}
\usepackage{array,dcolumn}
\usepackage{psfrag}
\usepackage{graphicx}
\usepackage{psfrag}
\usepackage{amsmath}
\usepackage{amsfonts}
\usepackage{amssymb}
\usepackage{accents}
\usepackage{pstricks,exscale}
\usepackage{verbatim}  
\usepackage{epstopdf}
\usepackage{graphicx}
\usepackage[caption=false]{subfig}
\usepackage{hyperref}

\usepackage{floatrow}
\newcolumntype{C}[1]{>{\let\newline\\\arraybackslash\hspace{0pt}}m{#1}}
\usepackage{color}
\usepackage[normalem]{ulem} 

\journal{Electrical Power and Energy Systems}

\begin{document}

\begin{frontmatter}

\title{A Strategy for Power System Stability Improvement via Controlled Charge/Discharge of Plug-in Electric Vehicles}

\author{Andrej Gajduk$^1$}
\author{Vladimir Zdraveski$^{1,2}$}
\author{Lasko Basnarkov$^{1,2}$}
\author{Mirko Todorovski$^3$}
\author{Ljupco Kocarev$^{*,1,2,4}$}
\cortext[cor1]{Author to whom any correspondence should be addressed.\\e-mail address: lkocarev@ucsd.edu\\postal address: Bul. Krste Misirkov, 2, P.O. Box 428 1000 Skopje, Macedonia\\telephone: +38923235400}

\address{$^1$Macedonian Academy of Sciences and Arts, Skopje, Macedonia}
\address{$^2$Faculty of Computer Sciences and Engineering, University ``Ss Cyril and Methodius'', Skopje, Macedonia}
\address{$^3$Faculty of Electrical Engineering and Information Technologies, University ``Ss Cyril and Methodius'', Skopje, Macedonia}
\address{$^4$BioCircuits Institute, UC San Diego, La Jolla, CA 92093-0402, USA}

\begin{abstract}
Plug-in electrical vehicles (PEV) are capable of both
grid-to-vehicle (G2V) and vehicle-to-grid (V2G) power transfer. The advantages of developing V2G include
an additional revenue stream for cleaner vehicles, increased stability and
reliability of the electric grid, lower electric system costs, and eventually,
inexpensive storage and backup for renewable electricity. Here we show how smart
control of PEVs can improve the stability of power grids using only local
frequency measurements. We evaluate the proposed control strategy on the IEEE
Case 3 and the IEEE New England power systems. The results show that V2G leads
to improved steady-state stability, larger region of stability, reduced
frequency and voltage fluctuations during transients and longer critical clearing times.
\end{abstract}


\end{frontmatter}

\section{Introduction}
\label{intro}

Global climate change, rising energy costs and limited fossil fuels resources have triggered a shift towards sustainable personal transport solutions with the plug-in electric car (PEV) playing the key role~\cite{penna2014climate,}. Several challenges still limit the adoption of PEVs are high initial vehicle cost; consumer acceptance; and infrastructure. Still, according to a recent forecast  PEVs  will make up around 1\% of the world vehicle fleet by 2020~\cite{duan2014forecasting}.

Currently, most PEVs establish connection to the grid only to recharge their batteries. But PEVs can also behave as distributed energy generators in the so called vehicle-to-grid (V2G) mode~\cite{kempton2001vehicle}. The V2G concept is based on the observation that PEVs are only in use or charging 2 hours a day~\cite{pearre2011electric} making them potentially available the rest 91\% of the time for other purposes. Researchers are exploring the capabilities of electric vehicles for peak power supply~\cite{drude2014photovoltaics,pang2012bevs}, renewable energy integration~\cite{han2014development,mwasilu2014electric}, regulation support~\cite{sortomme2012intelligent,liu2013decentralized} and spinning reserve~\cite{ota2012autonomous,ehsani2012vehicle}. Another group of research addresses novel applications such as reactive power compensation~\cite{deilami2011real} and current harmonic filtering~\cite{sun2014unified}. 

The power grid is one of the most complex systems that mankind has engineered. Because the grid has no storage, electricity production must be continuously managed to meet the fluctuating demand. If this balance between generation and consumption is not maintained at all times the system variables will start to drift from their nominal values. The ability to maintain frequency and voltage 
at their scheduled voltage is refereed to as stability of the power system. Small disturbances such as incremental load changes are area of interest for \textit{steady-state stability}. On the other hand, \textit{transient stability} deals with large disturbances such as short-circuits and line trips.

Scientists and engineers are constantly looking for new ways to improve the power grid stability using the concept of smart grid: smart control~\cite{rueda2013hybrid,sofla2012control}, smart demand response~\cite{schafer2015decentral}, smart managament of HVDC lines~\cite{fuchs2013placement} and smart topology control~\cite{bruno2012controlling}. Recently we have shown that transient stability can be improved with smart control of PEVs~\cite{gajduk2014improving}.
The main issue with most smart approaches is the reliance on a centralized entity that manages and controls a series of devices~\cite{richardson2013electric}. This is also true for most PEV applications that also propose aggregating a group of vehicles to facilitate management, which adds yet another layer of security and privacy concerns, latency related problems and high initial cost.

%
Frequency is an important characteristic of the current state of a given power.
A shortage of power is manifested as a decrease in frequency, conversely increased frequency indicates that there is excess generation in the system.
Any frequency based control system will thus act to reduce consumption and increase generation if the frequency starts to decrease and vice verse.
We use the readily available frequency measurements to control a fleet of PEVs connected at the distribution level in order to improve the overall power system stability. This approach does not require any complex communication infrastructure and expensive metering equipment unlike most smart solutions. In fact, measuring the frequency at your home or office can be done using a small PIC microcontroller~\cite{alam2013measurement}, that can also be used to govern the power exchange between the electric vehicle and the grid.

In this paper we suggest smart control strategy of PEVs that can improve the stability of power grids using only local frequency measurements. The strategy is evaluated on the IEEE Case 3 and the IEEE New England power systems. The results show that V2G leads to improved steady-state stability, larger region of stability, reduced frequency and voltage fluctuations and longer critical clearing times. This is the outline of the paper. 
First, in Sect.~\ref{sec:local_control} we propose a control strategy for the individual electric vehicles and develop a hardware solution for vehicle-to-grid. Next, in Sect.~\ref{results} we evaluate this control strategy using a structure preserving power system model, that aggregates the effect of PEVs connected at the same substation. For two test power systems we show that PEVs (1) improve the steady-state stability and robustness of a power system and (2) also result in better transient stability. Section~\ref{concl} concludes this paper.

\section{Local control}
\label{sec:local_control}

\subsection{Local control strategy} 

In power systems the balance between generation and load is kept by adjusting
the power output of the generators, which is known as automatic generation
control (AGC). In this process we distinguish two classes of generators: a) fast
responsive and b) slow responsive. Generators in hydropower plants and in
thermal power plants with gas turbines belong to the first class and are
highly valuable in AGC during disturbances in power systems. All conventional
thermal power plants, i.e. their generators, are not suited for tackling
large disturbances since their typical response times are in the range of
minutes. Usually, short reaction times in order of seconds are required in order
to bring back the power system into equilibrium as soon as possible and avoid
system-wide black-out \cite{machowski2011power}.

Besides the conventional power plants and renewable generation PEVs are another
possibility for power balancing in AGC during disturbances. Having
in mind their large number, which is forecasted to grow, it is expected that many
of them will be connected to the grid in the instance with a disturbance and
their presence and stored energy can be utilized to help in system
stabilization. At this point we would like to stress the fact that PEVs are very
flexible and may react in milliseconds exchanging power with the grid (in both
directions) through fast power electronic interfaces.

In \cite{gajduk2014improving} we proposed a novel control strategy that regulates power
exchange between PEVs and the power grid, based on the average turbine speeds at
conventional generators, in an effort to reduce the effects of large
disturbances. We model PEVs as loads, which they really are, but in certain
moments their power may become negative and they will be perceived by the system
as generators. A decrease in average turbine speeds signals a power shortage in
the system, thus PEVs are instructed to feed additional power to the grid, in
essence acting as small generators and vice versa. It was shown that by
regulating the power output of the PEVs, speed and voltage fluctuations during
disturbances can be significantly reduced. Furthermore, the critical clearing
time, that is the time to clear the fault which caused the disturbance, can be
extended by 20 to 40\%. This in turn yields more robust power system.

In this paper we go a step further and change the PEV control strategy by
proposing to use the local frequency measurements as a control signal for
adjustment of power exchange between individual electric-vehicle and the grid.

\begin{figure}[!htb]
\centering
\includegraphics[scale=1]{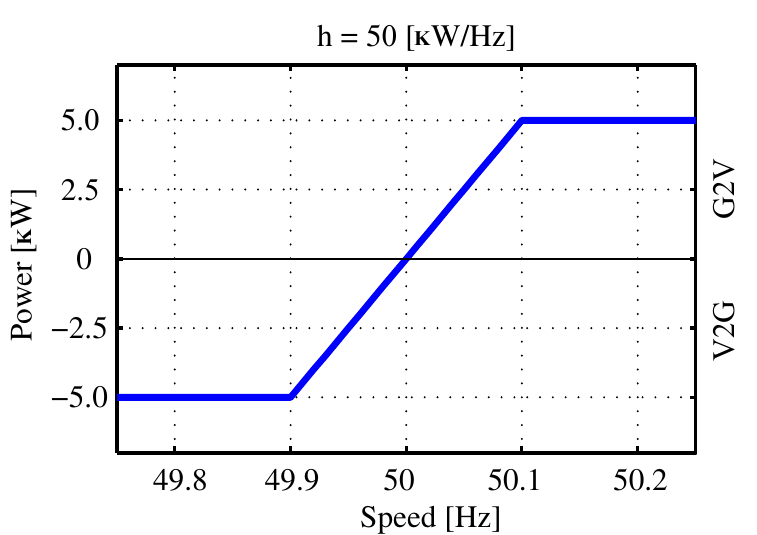}
\caption{The linear control strategy that
allows the PEV to behave as a generator ($f < 50 Hz$) or a load ($f > 50 Hz$).}
\label{fig:control_plot}
\end{figure}

We opted for a simple linear control (similar to turbine governors), where PEV
power output is linearly dependent on frequency deviation, with ramp limits of
$\pm 5$ kW at $\pm100$ mHz (Fig.~\ref{fig:control_plot})
\begin{equation}
P = \left\{
\begin{aligned}
-5~&{\rm kW}, \quad {\rm if~} \Delta f \le -100 {\rm ~mHz}\\
h \cdot \Delta f~&{\rm kW}, \quad {\rm if~} -100 {\rm ~mHz~} < \Delta f \le 100 {\rm ~mHz}\\
5~&{\rm kW}, \quad {\rm if~} \Delta f > 100 {\rm ~mHz}\\
\end{aligned}
\right.
\label{eq:p_control}
\end{equation}
where $h = 50 \mbox{ kW/Hz}$ is a control parameter and $\Delta f$ is the
frequency deviation from its nominal value (expressed in Hz).
According to this type of control when the frequency is lower than nominal 50 Hz
the PEVs start feeding power back to the grid (negative load). On the other
hand, if the frequency becomes higher than nominal 50~Hz 
PEVs are instructed to act as loads. Due to limitations of the connecting infrastructure, we take that the power exchange between any PEV and the grid can be maximum 5
kW~\cite{kempton2005vehicle1}.

Power frequency is never constant and even under normal operation it
fluctuates around its nominal value due to the power demand fluctuations, which
is a well know phenomena in the field of steady state stability
\cite{machowski2011power}. For this reason we suggest to use $d(\Delta f)/dt$,
i.e. the first derivative of $\Delta f$, as an indication that there is severe
disturbance in the system. A disturbance which causes frequency deviation with a
rate of 0.1 Hz/s or more acts as a trigger signal that enforces PEVs to follow
the proposed control strategy~(\ref{eq:p_control}). Another trigger signal is generated when the
disturbance subsides, that is in the moment when both $\Delta f$ and $d(\Delta
f)/dt$ are below a certain threshold (typically $\Delta f < 0.01$ Hz and
$d(\Delta f)/dt = 0.1$ Hz/s). The second trigger signal puts the PEVs into
``sleep mode'' and they stop following~(\ref{eq:p_control}).

\subsection{Hardware Prototype}
\label{hardware}

\begin{figure}[!htb]
\centering
\includegraphics[scale=.65]{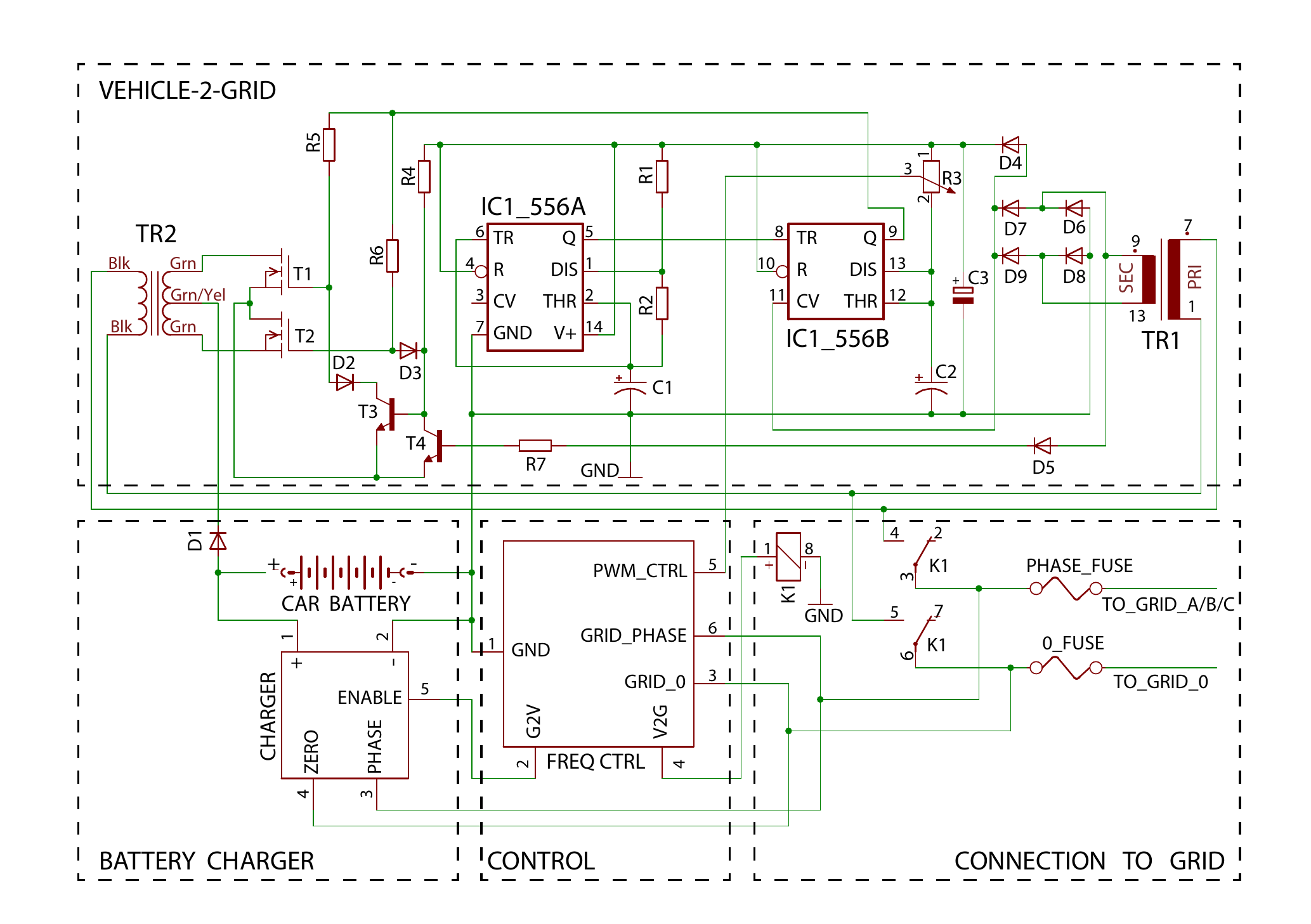}
\caption{A detailed schematic of the V2G module (top), the controller (middle),
the GRID connector relays (right) and the battery charger (left).}
\label{fig:hw_schematic}
\end{figure}

We propose a V2G extended battery charger
(Fig.~\ref{fig:hw_schematic}), that consists of three main
blocks:
(1) a battery charger, (2) an inverter circuit and (3) a controller that adjust inverter power to the frequency deviation, as described above.

The controller block implementation may vary from a very simple frequency
counter (such as ICM7216\footnote{\href{http://www.intersil.com/content/dam/Intersil/documents/icm7/icm7216b-16d.pdf}{\textcolor{blue}{http://www.intersil.com/content/dam/Intersil/documents/icm7/icm7216b-16d.pdf}}}) combined with a custom finite state machine, up to a powerful smart general
purpose computer, that offers a web configuration and control interface.
The controller block will continuously measure the frequency deviation and
switch on/off the charger (via G2V signal) or the inverter (via V2G signal). It will also regulate the inverter's output power based on the measured frequency deviation.

The inverter circuit is designed using IC556 (two IC555, placed on a single chip\footnote{\href{http://homemadecircuitsandschematics.blogspot.com/2012/10/designing-grid-tie-inverter-circuit.html}{\textcolor{blue}{http://homemadecircuitsandschematics.blogspot.com/2012/10/designing-grid-tie-inverter-circuit.html}}}) 
as pulse-width modulation (PWM) signal generator. It is a grid-tie inverter
with the transformer TR1 (top-right in Fig.~\ref{fig:hw_schematic}) as its
main supply and a referent sine wave generator. Inverter's power
depends on the PWM frequency and its high level duration, which is regulated with the resistance of R3. The generated power is returned to the GRID via the transformer TR2.
The physical characteristics of the resistors,
capacitors, diodes and MOSFET's as well as the PWM signal control itself may vary with the required power in a specific V2G charger.

\section{Stability analysis}
\label{results}

\subsection{System model}
\label{steady_state}

The dynamic behavior of a power system is described by a set of
first order differential equations with generators' rotor angle and speed taken
as state variables. With these equations, for each generator, we are modeling
the mechanical oscillations due to the power imbalance defined as a difference
between generator input and output. Generator input is equal to the turbine
mechanical power, while generator output depends on the conditions in the
transmission network and it is equal to the electric power injected in the
network. The latter, depends on the characteristics of the network, voltage
levels, load demand and generator electrical characteristics (transient
reactance and electromotive force).

Transient reactances and electromotive forces of generators are included in the
network model. All network elements (transmission lines and transformers) are
modelled by their $\pi$-equivalents circuits, while consumers are modelled by
power injection which are replaced by a constant nodal shunt admittance.
The resulting network model is described by the bus admittance matrix, while the
relation between voltages and currents is given by the nodal admittance
equation.

Consider a power system with $N$ buses of which $n$ are generator buses and $m$
are load buses. We assume that every load bus also has some PEVs connected to it
and aggregate them together in a PEV group (PEVG) to facilitate modeling.
Each load is modeled as a passive admittance
$${\underline{Y}_{i0} = \left( P_{i}+\jmath Q_{i} \right) / V_{i}^2 = G_{i0}+\jmath B_{i0}}$$. The power equations at each load bus are
\begin{align}
0 &= V_i^2 G_{i0} + \sum_k^N V_i V_k \left( G_{ik}\cos(\delta_i-\delta_k)+B_{ik}\sin(\delta_i-\delta_k) \right) + P_i^{PEVG} \nonumber\\
0 &= -V_i^2 B_{i0} + \sum_k^N V_i V_k \left( G_{ik}\sin(\delta_i-\delta_k)-B_{ik}\cos(\delta_i-\delta_k) \right) \label{eq:PQ_load}
\end{align}
where $G_{ik}$ and $B_{ik}$ are real and
imaginary part of the corresponding element of bus admittance matrix,
$\underline{Y}_{ik} = G_{ik} + \jmath B_{ik}$; $V_i$ and $\delta_i$ are the
voltage magnitude and angle at $i_{th}$ bus $\underline{V_i} = V_i
e^{\jmath \delta_i}$; and $P_i^{PEV}$ is the injected 
active power by PEVs at $i_{th}$ bus.

The generators are modeled as a constant electromotive force behind a transient
reactanse, and their dynamics are described by the swing equation

\begin{align}
\dot{\delta_i} &= \omega_i  \nonumber\\
\dot{\omega_i} &= \frac{1}{M_i} \left[ -D_i \omega_i + P_i^m - \sum_k^N V_i V_k \left( G_{ik}\cos(\delta_i-\delta_k)+B_{ik}\sin(\delta_i-\delta_k) \right)\right] \label{eq:swing}
\end{align}

\noindent where $\delta_i$ is the generator angle; $\omega_i$ is the generator speed; $M_i$ is the rotor inertia constant; $D_i$ is the damping coefficient; and $P_i^m$ is the mechanical power driving the generator turbine. In a sense this model resembles the ``structure preserving model''~\cite{van1985structure}, with simple passive loads and an added term for PEVs.

The power $P^{PEVG}_i$ is the aggregate effect of all electric vehicles at $i_{th}$ bus and is analogous to the power exchange with a single vehicle and also peaks when the frequency deviation reaches $\pm 100~{\rm mHz}$

\begin{equation*}
P_i^{PEVG} = \left\{
\begin{aligned}
-0.1 \cdot h_i~&{\rm MW}, \quad {\rm if~} \Delta f \le -0.1 {\rm ~Hz}\\
\Delta f_i \cdot h_i~&{\rm MW}, \quad {\rm if~} -0.1 {\rm ~Hz~} < \Delta f \le 0.1 {\rm ~Hz}\\
0.1 \cdot h_i~&{\rm MW}, \quad {\rm if~} \Delta f > 0.1 {\rm ~Hz}\\
\end{aligned}
\right.
\end{equation*}

\noindent where $h_i$ is a control parameter and $\Delta f_i = \Delta \omega_i /
2 \pi  $ is the frequency deviation at $i_{th}$ bus. 
In order to reduce the number of control parameters we take them to be
proportional to the load consumption at that bus \textit{i.e.} 
$$
h_i = h \frac{P_i}{\sum_{k=1}^N P_k}
$$
Our goal is to analyze how different values for the control parameter $h$ affect the system stability.







\begin{figure}[!htb]

\centering
\begin{tabular}{c c}
\subfloat[]{\includegraphics[scale=0.315]{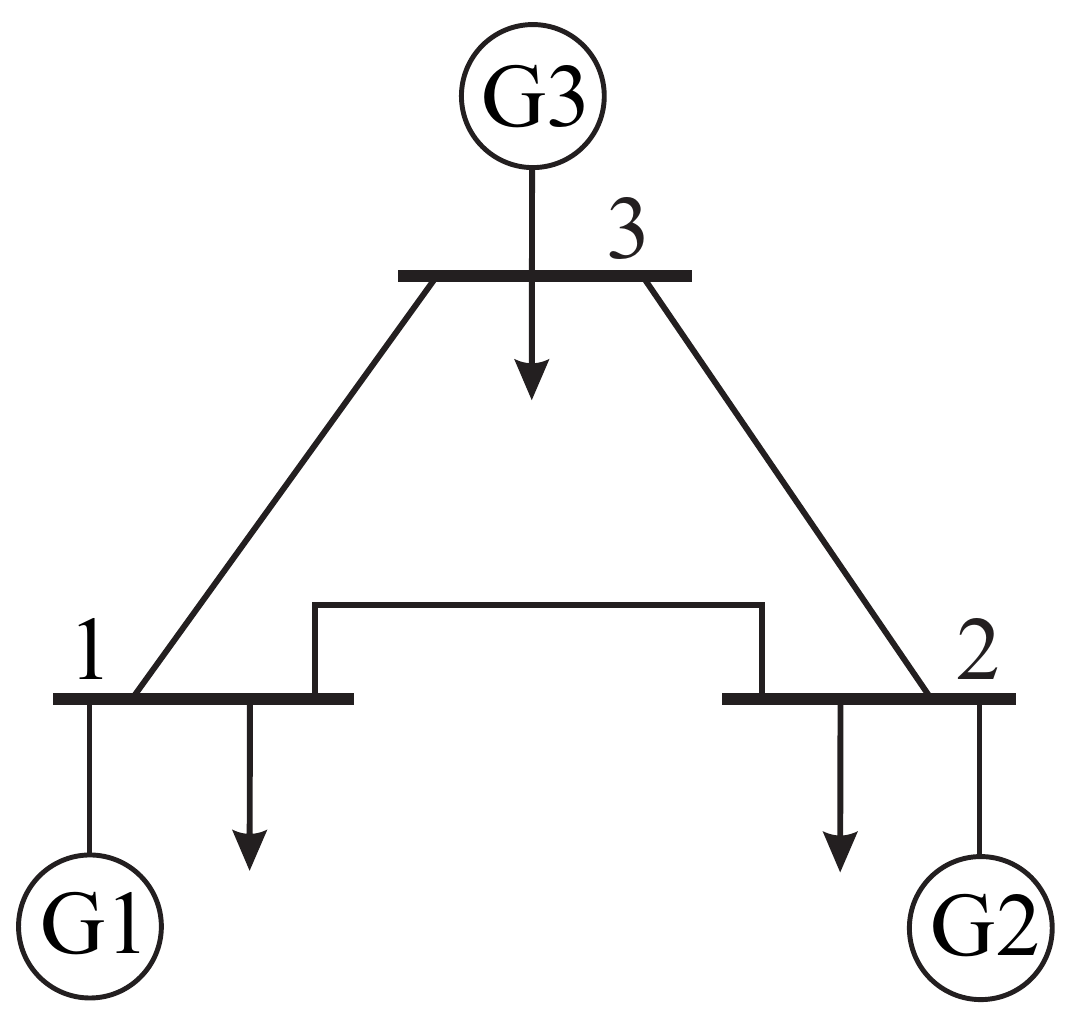}}
&
\subfloat[]{\includegraphics[scale=0.166]{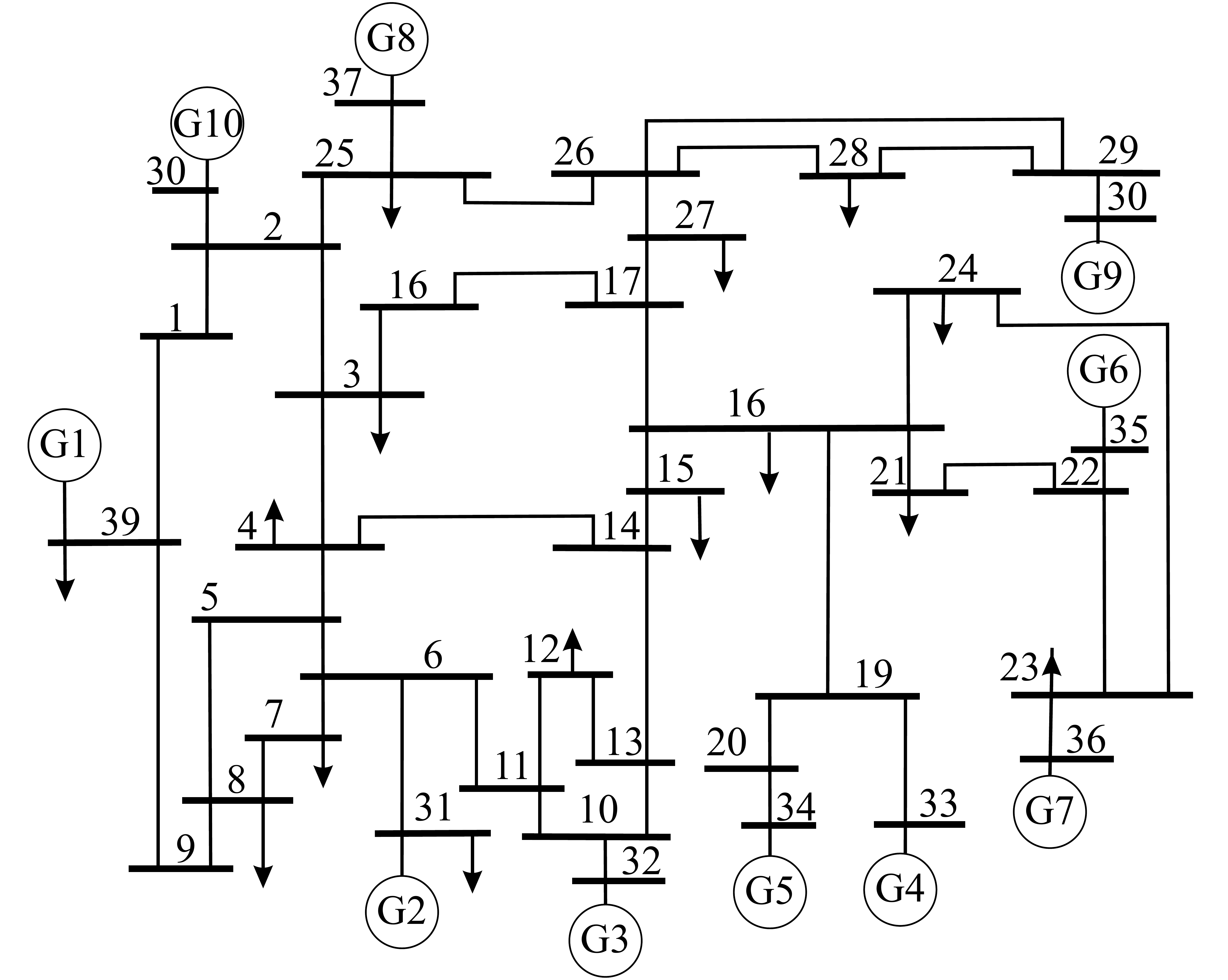}}
\end{tabular}

\caption{
Sample power systems for evaluating the control strategy (a) IEEE Case 3
\cite{athay1979practical} (b) IEEE New England \cite{pai1989energy}. }
\label{fig:sample_grids}

\end{figure}

The proposed control strategy was tested in two test systems
(Fig.~\ref{fig:sample_grids}). The first one is a 3-bus 3-machine system, while
the second one is 39-bus 10-machine system (New England). In particular, we
investigated how PEVs can improve the stability in power systems and the results
are given in the following two sections. 
The software used for dynamical simulation of PEVs is available online as open source~\footnote{\href{https://github.com/gajduk/vehicle2grid-dynamic-simulation-PSSE}{\textcolor{blue}{https://github.com/gajduk/vehicle2grid-dynamic-simulation-PSSE}}}.

\subsection{Steady-state stability}

A power system is steady state stable if it is able to reach a new
stable configuration following a small disturbance in the system. Such
disturbances are continuously present in normal system operation and they
include load fluctuation, actions of automatic voltage regulators or switching
operations of less important system elements. In other words steady state
stability applies to system events under which there are very gradual and
infinitesimally small power changes. The new stable state is very close to the
pre-disturbance operating point. In such cases the equations describing the
power system dynamics may be linearized for analytical purposes.

It is well known that a system is steady-state stable if the Jacobian matrix,
obtained in the process of equation linearization using Taylor series, has only
eigenvalues with negative real parts \cite{sauer1990power}. Furthermore, the
largest real part of any eigenvalue, denoted with $\alpha$, of the Jacobian
matrix also serves as an indicator of the overall system
stability~\cite{doyle1982analysis,becker1994robust}. Therefore, we use it to
compare the stability of power systems with and without V2G.

\begin{figure}[!htb]

\centering
\begin{tabular}{c c}
\subfloat[Steady state stability of the New England system]{\includegraphics[scale=0.75]{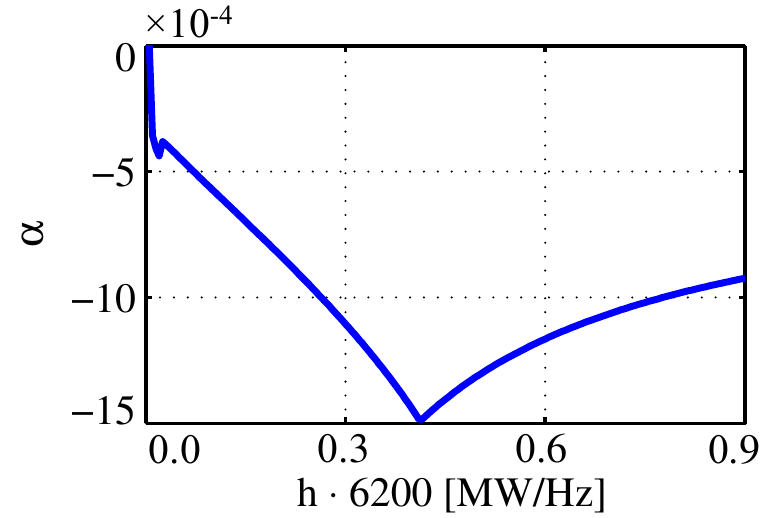}}
&
\subfloat[Steady state stability of the case 3 system]{\includegraphics[scale=0.75]{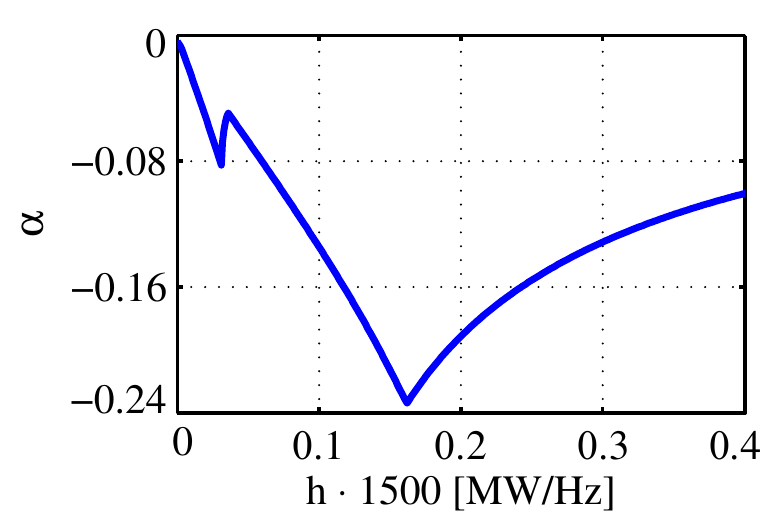}}

\end{tabular}

\caption{
Stability indicator $\alpha$ as a function of the control parameter $h$.
Minimums are observed at $h = 0.4$ for IEEE New England and at $h = 0.17$ for
IEEE Case 3. The control parameter values are scaled proportionally to the total consumption of the power system (6200 MW New England and 1500 MW for case 3).}
\label{fig:ss_stab}
\end{figure}

The results have shown that the steady-state stability of power systems improves
when PEVs, which is shown as a decrease in $\alpha$ in Fig.~\ref{fig:ss_stab}.
This improvement in stability increases proportionally with the
control parameter $h$ up to a certain point and then starts to decrease.
In our opinion this is due to the PEVs
overreacting to the disturbance, which negates the benefits of V2G. If $h$
is very large there is a possibility that the system will become less stable.
Therefore, determining the critical value of $h$, which yields minimum
value of $\alpha$ is an important engineering challenge.

\subsection{Transient stability}
\vspace{10px}

Power systems may experience severe disturbances which include:
short circuits with or without significant network topology change, switching
operations of important lines/transformers and sudden application or removal of
big load. Transient stability of a power system refers to the ability of the
system to reach a stable condition following any large disturbance in the
transmission network. In these cases the right hand sides of (\ref{eq:PQ_load})
and (\ref{eq:swing}) undergo significant changes for two reasons: 1) there are
large excursions of generator rotor angles and the power-angle relationship has
to be taken in its original nonlinear form, 2) there are large changes in the
coefficients $G_{ij}$ and $B_{ij}$ due to bus admittance matrix changes which
are substantial. In these situations the new stable state may be very different
from the pre-disturbance operating state and equation linearization is not
applicable  \cite{machowski2011power}. In transient simulations we identify
three stages: 1) pre-fault stage, 2) fault stage and 3) post-fault stage.

During the pre-fault stage the system is in equilibrium and the state variables
are constant. Then at $t = t_0$ a fault occurs that changes the system topology,
thus changing the admittance matrix which in turn cancels the
balance between consumption and production the system state variables begin to
deviate. At $t = t_{cl}$ the fault is cleared and the
admittance matrix returns to its pre-fault state. The initial conditions for the
post fault stage are the system parameters at $t_{cl}$. Depending on the these
initial values the system will either return to stable operation or lose
synchronization.

In transient stability analysis it is important to determine if the system will
return to stable operation after a fault has been cleared. This is greatly
facilitated if one knows the region of asymptotical stability (RAS) which is
defined as the largest region in parameter space for which the system state
converges to equilibrium. If we know the RAS then we can determine whether the
system will remain stable for a given initial conditions just by checking if
those conditions lie inside the specified region.
Consequently, a larger RAS means that the system is more stable and can handle
more severe disturbances.

\begin{figure}[!htb]
\includegraphics[scale=.5]{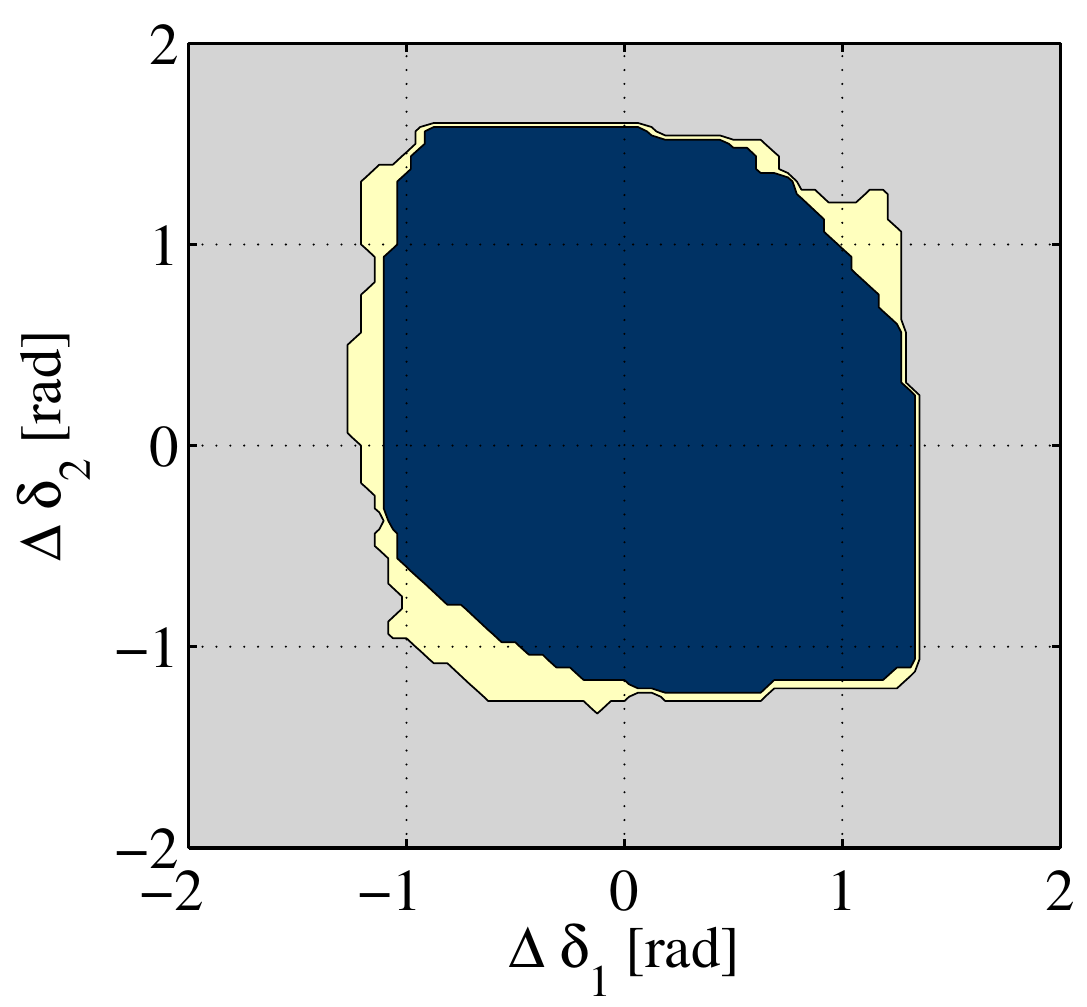}
\caption{Region of asymptotical stability (RAS) for IEEE Case 3 power system. The stable region for the standard power system (blue) is expanded with vehicle-to-grid (yellow).}
\label{fig:RAS}
\end{figure}

\begin{figure}[!htb]
\CenterFloatBoxes
\begin{floatrow}
\ffigbox
  {\includegraphics[scale=.83]{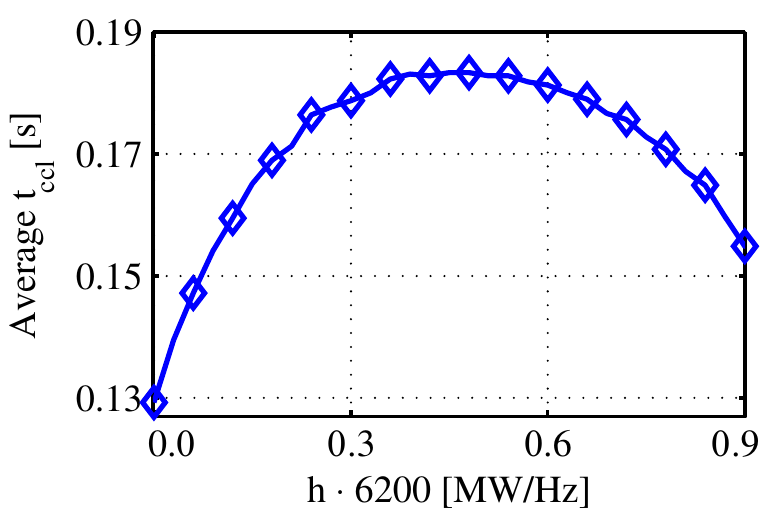}}
  {\caption{The critical clearing time $t_{ccl}$ for three-phase short circuits averaged for all buses}\label{fig:psse_t_ccl}}
\killfloatstyle
\ttabbox
  {\begin{tabular}{ | c ||  c |  c |  c |  c | } 
 \hline 
Bus \textbackslash h & $0$ & $.3$ & $.6$ & $.9$  \\ \hline 
7 & .13 & .14 & .15 & .16    \\ \hline 
12 & .23 & .25 & .26 & .27   \\ \hline 
13 & .12 & .13 & .14 & .14    \\ \hline 
18 & .10 & .11 & .12 & .12    \\ \hline 
25 & .12 & .13 & .14 & .15    \\ \hline 
32 & .15 & .16 & .17 & .18    \\ \hline 
38 & .10 & .11 & .11 & .11    \\ \hline 
\end{tabular}
  }
  {\caption{Critical clearing times for different bus faults and various control constants for the case 39.}\label{tab:psse_t_ccl}}
\end{floatrow}
\end{figure}

Interestingly, the RAS expands when V2G is implemented for power regulation
(Fig.~\ref{fig:RAS}). The RAS is only drawn for the IEEE Case 3 since in this
case it is a two-dimensional figure, while for the IEEE New England it is rather
difficult to visualize it. However, the results have shown improvement in
stability, i.e. longer critical fault clearing times, for IEEE New England as
well (Table \ref{tab:psse_t_ccl}).
\renewcommand{\arraystretch}{1.15}

The critical clearing
time $t_{ccl}$ is defined as a time instance after the fault occurrence in which
the fault should be cleared in order to have stable operation afterwards. This
means that the fault clearing time $t_{cl}$ should be $t_{cl} \leq t_{ccl}$ for
stable operation. In case when $t_{cl} > t_{ccl}$ the stable operation will be
lost. The critical clearing time is a very important power system
characteristic and can be used to asses the system stability when computing the
RAS is unfeasible. Again the simulations show that the system with V2G is more
robust to disturbances indicated by longer critical clearing times
(Fig.~\ref{fig:psse_t_ccl} and Table~\ref{tab:psse_t_ccl}).

\begin{figure}[!hb]

\begin{tabular}{c c c}
\subfloat[]{\includegraphics[scale=.81]{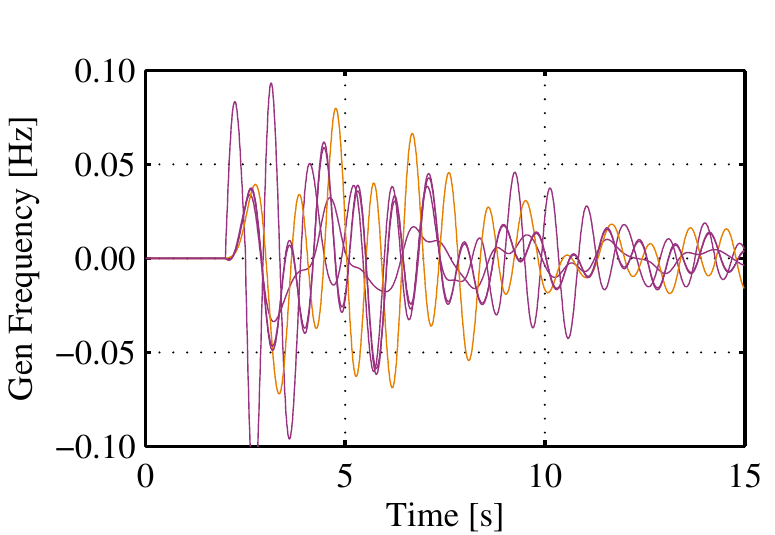}}
&
\subfloat[]{\includegraphics[scale=.81]{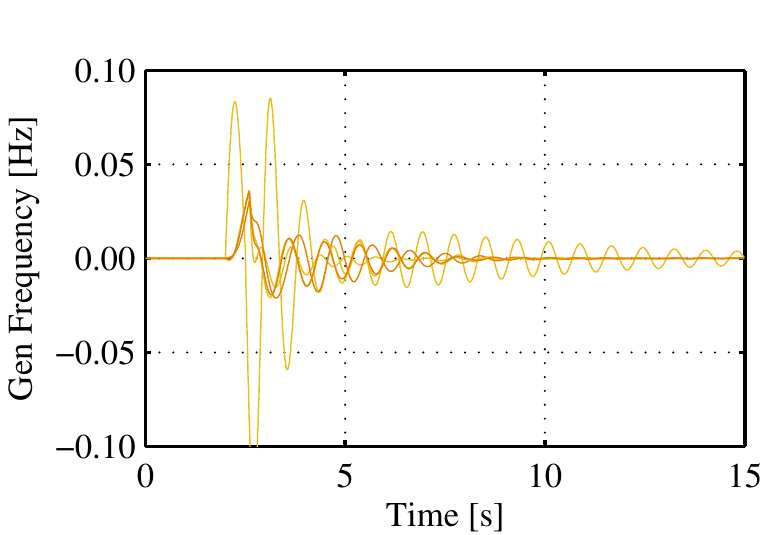}}

\\ \noindent

\subfloat[]{\includegraphics[scale=.81]{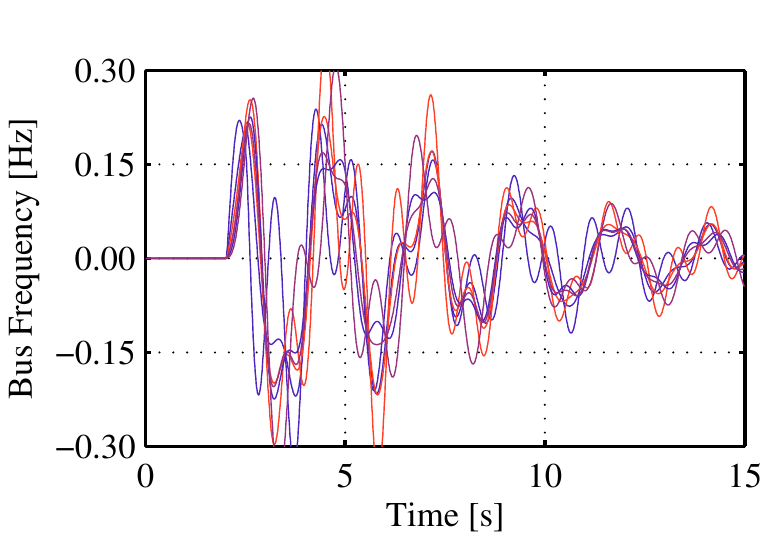}}
&
\subfloat[]{\includegraphics[scale=.81]{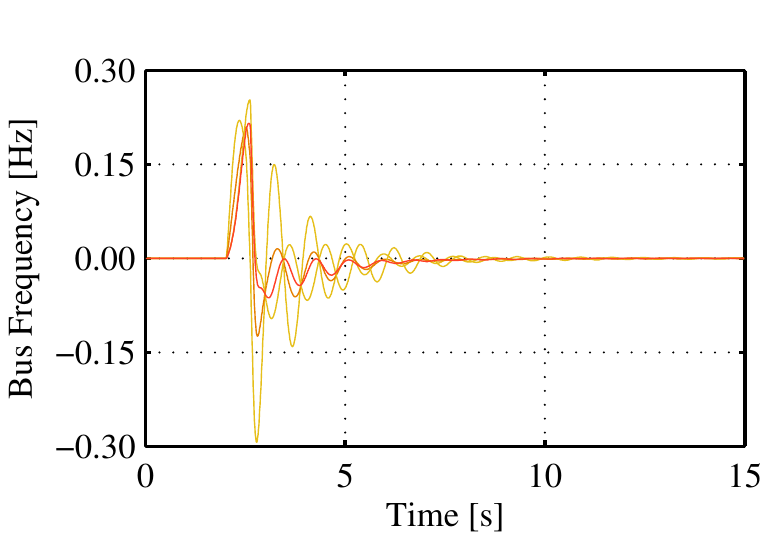}}

\\ \noindent

\subfloat[]{\includegraphics[scale=.81]{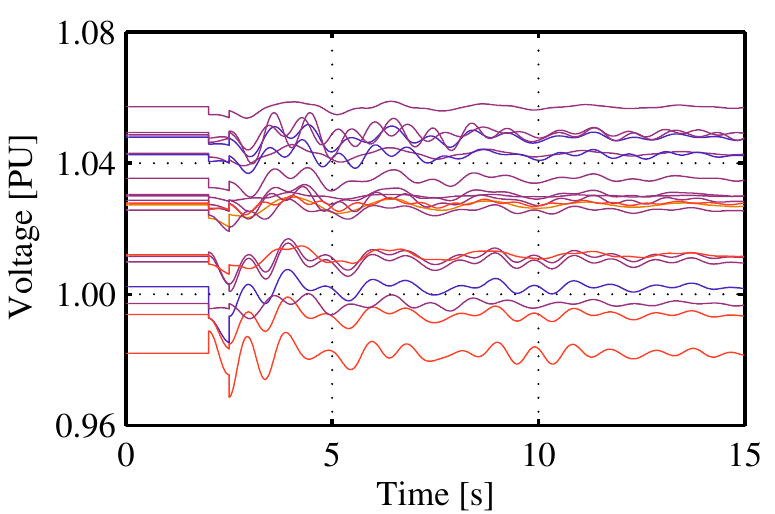}}
&
\subfloat[]{\includegraphics[scale=.81]{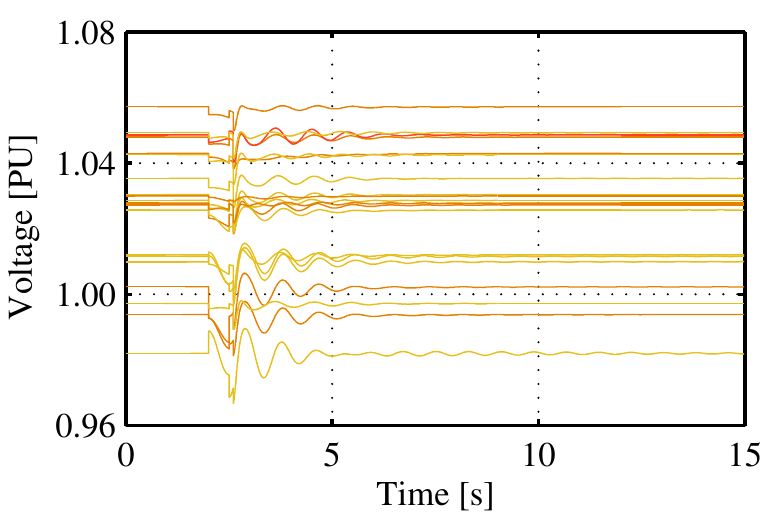}}

\end{tabular}

\caption{
Generator frequencies (a,b), bus frequencies (c,d) and system voltages (e,f) that result from a branch trip (5-6) that start at 2 [s] and lasts 0.5 [s]. The deviation are damped out faster if PEVs are used h= 5 * 6200 [MW/Hz] (b,d,f) as opposed to the standard system without PEVs (a,c,e).
}
\label{fig:fluctuations}
\end{figure}

During transients voltage and frequency start to deviate from their nominal
values which is dangerous and can damage sensitive 
appliances. Therefore it is imperative that they return to their prescribed
values as quickly as possible after a fault in the system occurs. As it turns
out, V2G not only helps to lengthen the critical clearing time, but also reduces
the voltage and frequency fluctuations in the post faults stage
(Fig.~\ref{fig:fluctuations}).

\section{Conclusions}
\label{concl}

In this paper we proposed a control strategy 
and developed a bidirectional charger (G2V and V2G) for plug-in
electric vehicles.
The control is performed locally using only local frequency measurements, thus avoiding the
need for a complex communication infrastructure and centralized control. In this
way, we eliminate concerns about the network latency, data privacy and security
and a whole range of other computer networks' related issues.  In addition, the
entire control system can be easily implemented using cheap and available
components.

We have demonstrated the applicability of the proposed control strategy using
two power systems from literature: IEEE Case 3 and IEEE New England. The power
system stability can be improved by smart management of plug-in electric
vehicles in both steady-state and transient stability. The system becomes more
robust with respect to small disturbances and the largest real part of any
eigenvalue decreases significantly: in the case of New England it goes from
$-0.1899 \cdot 10^{-2}$ to $-0.18 \cdot 10^{-4}$, which is reduction of two
orders of magnitude. When large disturbances occur the system parameters like
voltage and frequency fluctuate less and it takes less time for the system to stabilize.
Furthermore, the critical clearing time for major faults is longer thus giving
more time for automatic prevention mechanisms to react to the fault.  We have
also shown that the RAS is larger when plug-in electric vehicles are
used.

There is a plenty of space for future work, which may be concentrated on many
technical and regulatory issues. In order to get more realistic representation
of the contribution of PEV in dynamic studies we may take into account PEV
availability considering some probabilistic parameterss such as driving habits
and vehicle traffic simulation. Another possibility might include different
control strategies with vehicle power output being non-linearly dependent on
frequency deviation. And last, but not least PEV owners should be given
incentives for participation in dynamic stability improvement which should
probably require regulatory changes in the grid codes for transmission system
operation.

\end{document}